\documentclass[twocolumn,prb,showpacs,amsmath,a4paper,floatfix,preprintnumbers]{revtex4}

\usepackage{graphicx}
\usepackage{dcolumn}
\usepackage{bm}

\begin{document}

\title{O adsorption and incipient oxidation of the Mg(0001) surface}

\author{Elsebeth Schr{\"o}der$^1$}
\author{Roman Fasel$^2$}
\author{Adam Kiejna$^3$}
\affiliation{$^1$Department of Applied Physics, Chalmers University of
Technology and G\"oteborg University, SE--41296 G\"oteborg, Sweden}
\affiliation{$^2$Swiss Federal Laboratories for Materials Testing
and Research (EMPA), nanotech@surfaces Laboratory, Feuerwerkerstrasse 39, 
CH--3602 Thun, Switzerland}
\affiliation{$^3$Institute of Experimental Physics, University of Wroc{\l}aw, 
Plac M.\ Borna 9, PL--50--204 Wroc{\l}aw, Poland}

\preprint{Applied Physics Report no.\ 2001--67}

\date{June 27, 2003} 

\begin{abstract}  
First principles density functional calculations are used to study the early 
oxidation stages of the Mg(0001) surface for oxygen coverages 
$1/16\le\Theta \le 3$ monolayers. It is found that at very low coverages 
O is incorporated below the topmost Mg layer in tetrahedral sites.
At higher oxygen-load the binding in on-surface sites is increased but at one 
monolayer coverage the on-surface binding is still about 60 meV weaker 
than for subsurface sites. 
The subsurface octahedral sites 
are found to be unfavorable compared to subsurface tetrahedral sites and to 
on-surface sites. At higher coverages oxygen adsorbs both under the surface 
and up. Our calculations predict island formation and clustering of 
incorporated and adsorbed oxygen in agreement with previous calculations.
The calculated configurations are compared with the angle-scanned
x-ray photoelectron diffraction experiment to determine the geometrical
structure of the oxidized Mg(0001) surface.
\end{abstract}  


\maketitle

\section{INTRODUCTION} 

The process of oxidation of metal surfaces is of considerable fundamental 
scientific interest as well of paramount technological 
importance.\cite{classical_oxide_ref}
Corrosion and passivation are just two examples of either destructive or
useful processes linked directly to this phenomenon, which are well-known
from everyday life.
Oxidation of metal surfaces begins with dissociative chemisorption of oxygen
on a clean surface and is followed by the formation of a film of metal oxide.
In-between these events there are many elementary processes.
The study of Al and Mg oxidation is particularly important
because Al and Mg 
belong to the group of so-called simple metals and are considered
model systems for studies of oxidation of transition metals.
Both metals exhibit high reactivity with oxygen and oxidize rapidly.
Aluminum oxides (alumina) form several different phases where the structure
of some of them was only recently identified by combined {\em ab initio\/} 
density functional theory (DFT) and
experimental studies.\cite{Yourd99} Magnesium oxide is known to 
eventually
form crystals of the rock-salt structure,\cite{Nog}
but can also experience complex reconstructions at partial 
coverages.\cite{Else01,Goon02}
Both Mg and Al oxides play an important role in catalyst support and
find many other useful applications.

The number of experimental%
\cite{Goon02,NamDG81,Ghi81,Hay81,FloMa82,Thi89,Cron89,MitOCS98,Dri99,Lac94,Esa97}
studies and theoretical\cite{Else01,Bung97} treatments devoted to the
oxidation of magnesium is relatively limited compared to the more
extensively studied oxidation of aluminum.
In this paper we present a systematic study of the initial oxygen 
incorporation and island formation by first-principles theory
calculations and we use the results to interpret high-quality 
x-ray photo-electron diffraction (XPD) measurements
for low O$_2$ doses. 

The existing picture of the Mg(0001) surface oxidation processes dates
back to the early 1980's. Based on the extensive low-energy electron diffraction
(LEED), Auger electron spectroscopy, electron energy loss spectroscopy,
and work function measurements Namba et al.\ proposed~\cite{NamDG81}
a four-stage model consisting of: (1) dissociative oxygen chemisorption
at random sites followed by oxygen incorporation,
(2) assembly of incorporated oxygen atoms into islands and lateral growth,
(3) oxide formation from surface Mg atoms and subsurface O atoms which starts
at O$_2$ exposures around 2--3 L (Langmuir) or 0.5--1 monolayer of total
coverage, and (4) oxide thickening.
Several other experimental studies provided arguments for a three-stage
model of oxidation that was proposed at approximately the same time.
In this model \cite{FloMa82,Thi89} atomic oxygen is
directly incorporated into magnesium (below the top
Mg layer) right after on-surface dissociation. 
In the next step there is simultaneous formation of an oxide
layer and a decrease in the oxygen on the surface.\cite{Thi89}
Finally, there is oxide thickening and transformation into 
a rocksalt structure.

This model of immediate incorporation of oxygen gained support from
the x-ray photo-electron spectroscopy study of Ghijsen et al.,\cite{Ghi81}
and particularly from the measured sharp decrease in the work function of
the Mg(0001) surface upon initial O exposure.\cite{Hay81}
The formation of an O-($1\times 1$) subsurface layer up to monolayer
coverage and its subsequent transformation into epitaxial oxide was also
reported.\cite{Hay81}
The positions of the subsurface oxygen atoms were deduced to be octahedrally
coordinated where the oxygen has an environment similar to that in MgO.\cite{Thi89}
This is in agreement with a more direct experimental evidence for oxygen
incorporation in the octahedral interstitial sites of the first two interlayer
spacings of Mg during the initial stages of magnesium oxidation,\cite{Cron89}
and with later inelastic ion scattering experiments on a polycrystalline
Mg surface.\cite{Lac94,Esa97}
A very recent scanning tunneling microscopy (STM) study of the oxidation of
Mg(0001) suggests that at low oxygen exposures (up to 2 L) the incorporated
oxygen atoms form a single layer underneath the top layer of Mg.\cite{Goon02}
The idea of immediately populated subsurface sites seems also to be supported
by recent measurements by Mitrovic et al.,\cite{MitOCS98} however, they show
that most of the oxygen (90\%) remains over the surface and only a small
fraction goes below the surface. Thus, the question regarding the most
favored adsorption sites is still debated and requires further analysis.

On the theoretical side, the local density-functional theory (DFT) calculations 
of Bungaro et al.\cite{Bung97} show that oxygen is incorporated below the 
surface forming a $(1\times 1)$ subsurface lattice.
A recent DFT-based lattice gas model study\cite{Else01} of island formation in the 
early Mg(0001) oxidation stages
has shown that at very low filling the oxygen 
atoms adsorb in the top-most subsurface (tetrahedral) B-sites in an 
ABAB... stacking of magnesium.
At higher oxygen load also the deeper sites are filled, and
the adsorbed oxygen atoms form dense clusters 
just below the surface and in further subsurface 
locations.\cite{Else01,Bung97} For larger clusters 
the oxygen-oxygen interaction effects drive the oxygen
atoms further into the magnesium subsurface layers and bulk.

The rest of this paper is organized as follows.
In Section II systematic DFT calculations in the range of coverages 
between 1/16 and 3 monolayers (ML) are presented and used for
the study of the initial oxygen incorporation and island
formation.
The previous experimental and theory results are the starting base for our
first-principles calculations of the very beginning of O adsorption and oxide
formation at the Mg(0001) surface.
In Section III we describe our high-quality XPD measurements for
low O$_2$ doses corresponding approximately to global
coverages between 0.1 and 2 ML of atomic oxygen,
and in Section IV
our DFT results are used to interpret the experimental XPD data.

\section{THEORY}

The first-principles calculations are carried out using the plane-wave
density-functional \textsc{dacapo} code\cite{DACAPO}
with the generalized gradient
approximation (GGA) for the exchange-correlation energy functional\cite{Per92}
and with ultrasoft pseudopotentials\cite{Van90} to represent the ionic cores.
The clean Mg(0001) surface is modeled by periodic slabs consisting of six
to seven magnesium layers separated by 21 {\AA} of vacuum.
The plane-wave basis set with 25 Ry energy cutoff is used.
The O atoms are adsorbed on one side of the slab only and the
electric field arising due to the asymmetry of the system is compensated for
by a consistent dipole correction.\cite{NS92,Ben99}
A $4\times 4\times 1$ mesh of Monkhorst-Pack special $k$-points for the
$4\times 4$ atom surface unit cell and a Fermi-surface smearing of 0.2 eV
are applied to the Brillouin-zone integrations.
For smaller cells the number of k-points are increased accordingly up to a
$16\times 16\times 1$ mesh for the $1\times 1$ surface unit cell.
The positions of atoms in the three to four topmost magnesium layers,
and of all the
oxygen atoms, are fully optimized until the sum of the Hellmann-Feynman
forces on all unconstrained atoms converges to less than 0.05 eV/\AA.
The forces acting on the ions in the unit cell are derived from the converged
charge densities, and the atom dynamics is determined using a preconditioned 
quasi-Newton method based on the
Broyden-Fletcher-Goldfarb-Shanno algorithm.\cite{NumR}

The study involves very low coverages of oxygen, from one O atom
in a large surface cell up to coverages of three monolayers. We define
the coverage $\Theta$ as the ratio of the number of adsorbed atoms to
the number of atoms in an ideal substrate layer.
The binding energy per adsorbate is calculated relative to the energy of an
isolated, spin-polarized O$_2$ molecule and the clean, relaxed Mg-surface.

\subsection{The clean Mg(0001) surface}

Before studying the effect of oxygen chemisorption on the Mg(0001) surface,
we determined the bulk and bare-metal-surface structures. The
calculated lattice constants for hcp Mg is $a= 3.19$ \AA\ and $c/a= 1.64$,
which agree very well with the experimental values $3.21$ \AA\ and 1.623
(Ref.~\onlinecite{mglattconst}) and other GGA calculations.\cite{WaKi01}
The bulk modulus, ignoring lattice vibrations, is found to be
34.1 GPa,\cite{Ziam01} in good agreement with the measured value 35.5 GPa
(obtained from the elastic constants of Ref.~\onlinecite{mgbulkmod}) 
and with the GGA calculations of Ref.~\onlinecite{WaKi01}.

The relaxations of the surface interlayer spacing with respect
to the bulk spacing show 1.5\% expansion, 0.4\% contraction,
and 1.3\% contraction of the first, second and third interlayer distance,
respectively. The expansion of the first interlayer distance compares well
with the experimental\cite{DavHRP92} value of 1.7\%, and with other
GGA calculations.\cite{WaKi01}
The small contraction of the second interlayer spacing disagrees with results
of experiment and other calculations which predict a small expansion. Since the
magnitude of this relaxation is very small this discrepancy should not
have any significant effect on the results for O-adsorption.
The calculated work function is 3.72 eV for the relaxed surface and
agrees well with the experimental value 3.84 eV.~\cite{Mich77}

\subsection{Low coverages ($\Theta\leq 0.5$) of oxygen}

\begin{figure}
\scalebox{0.8}{\includegraphics{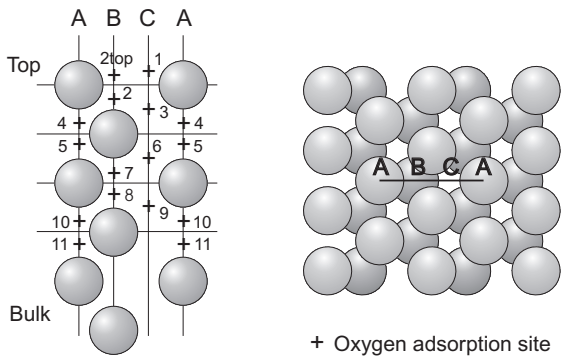}}
\caption{\label{figsketch}
Sketch of the adsorption sites of oxygen at low coverage.
The positions indicated by crosses to the left are the 
adsorption sites at 1/16 ML single 
O-atom adsorption.}
\end{figure}

Having obtained reliable results for the clean, relaxed Mg(0001) surface
we calculate the binding of oxygen in on-surface and subsurface sites.
The locations of the possible adsorption sites of Mg(0001) are sketched
with crosses in Fig.~\ref{figsketch}.
In our calculations the results for single or nearest-neighbor pairs
of O atoms adsorbed at the $4\times 4$, $2\times 2$, and $1\times 2$
surface cells were used as 1/16, 1/8, 1/4, and 1/2 ML coverage data.

\begin{figure}[b]
\scalebox{0.55}{\includegraphics{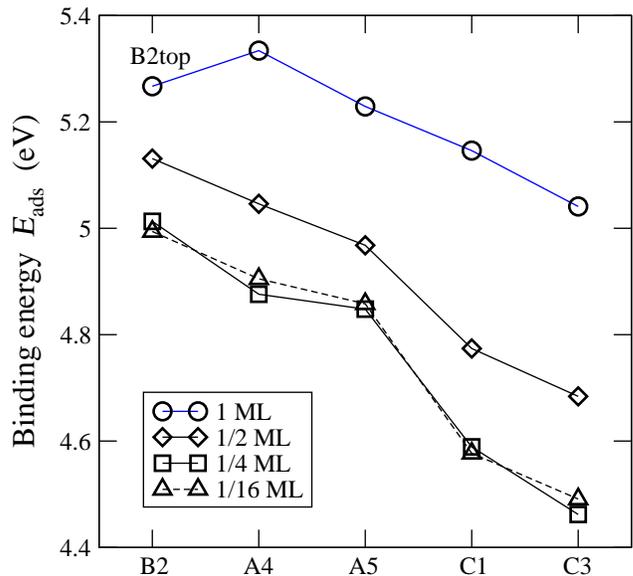}}
\caption{\label{f2}
The calculated low-coverage binding energies for different
adsorption sites as indicated in Fig.~\protect\ref{figsketch}.
For 1 ML coverage the B2 site is replaced by the slightly on-surface site B2top.
For coverages $\Theta \leq 0.5$ ML the only stable on-surface site is C1.}
\end{figure}

In Fig.~\ref{f2} we present the calculated low-coverage binding energies
with respect to the energy of an isolated O$_2$ molecule.
The adsorption of oxygen atoms on the Mg(0001) surface induces further
changes to the surface structure and the electronic properties.
However, the positional changes of the Mg atoms are in most
cases not very large. For the $2\times 2$ supercell with
a single O-atom
slightly below the surface (site B2 in Fig.~\ref{figsketch})
or in first-sublayer adsorption sites (sites A4 or A5)
the Mg atoms move less than 3\% vertically and 0.6\% laterally
from their clean-surface positions.
To estimate the sensitivity of the binding energy to
positional relaxations of the atoms we also calculated an even distribution
of four oxygen atoms in one of the tetrahedral subsurface sites (sketched
as site B2 in Fig.~\ref{figsketch}) in the $4\times 4$ system, and
compared the binding energy to the binding energy of one site-B2 atom
in the $2\times 2$ system.
The two configurations relax to slightly different atomic positions
(but both converged within the force requirement) and we found the difference
in binding energy per oxygen atom to be 5 meV. We therefore estimate the
relaxation-imposed inaccuracy of the binding energies to be approximately
5 meV.


\textit{On-surface adsorption}.
It has previously been suggested that for low coverages the
O atoms are preferentially adsorbed into \textit{on-surface\/}
sites,\cite{MitOCS98} as is the case for, e.g., oxygen adsorption on
Al(111),\cite{KiLu01} or that on-surface adsorption sites exist although 
having
lower binding energies than the lower-lying adsorption sites.\cite{Bung97}
For the Al(111) surface an additional O is incorporated only when 1 ML
of on-surface O is complete.\cite{KiLu01} A similar situation is found for
a transition metal, the Ru(0001) surface.\cite{Reuter}
For this reason on-surface adsorption was closely studied.
In summary, however, we find the initial oxygen adsorption
to be a subsurface process.

Of the on-surface adsorption sites the natural candidates are the C1 and B2top
hollows (Fig.~\ref{figsketch}). However, our calculated low-coverage binding
energies, plotted in Figure \ref{f2}, show that for single atom adsorption at
coverages $\Theta \le 0.5$ (corresponding to a single O atom in the
$4\times 4$, $2\times 2$, and $1\times 2$ surface unit cells) the only
stable on-surface adsorption site is the one denoted as C1,
and the binding energy of this site is
significantly less than that of any of the subsurface sites.
We find \textit{no\/} single-adsorbant energy minimum near
the position where we would expect the on-surface B2top-site.
Instead, when relaxing dilute systems with O in the B2top sites the 
surrounding Mg atoms move laterally by a small amount, enough
to let the O-atoms sink into the subsurface B2 (tetrahedral) site.
Only if a pair of O atoms are adsorbed as nearest neighbors,
or at increased O loading ($\Theta \geq 1$), can one of the O atoms be
stabilized on the surface in the B2top site, albeit at smaller binding energy
than in the slightly lower-lying B2-site.
Thus, for low coverages our DFT results point to
the C1 site as the only possible, but energetically unfavorable, on-surface
adsorption site. 

These results are in agreement with the theory results of Bungaro et 
al.\cite{Bung97}
as to subsurface adsorption sites being more favorable than on-surface
ones, but  contradict their prediction of a stable on-surface B2top-site
for low coverages. It should be noted, however, that for coverages
$\Theta < 1$ Bungaro et al.\cite{Bung97} relaxed only the O atoms but not
the Mg-lattice, and
that their $k$-point sampling was limited to the $\Gamma$ point only.
Thus our results underline the importance of relaxing the substrate.

\textit{The subsurface adsorption}.
By extensive searches for subsurface adsorption sites we find that
within the two top Mg-layers the tetrahedral sites (sites B2, A4, and A5)
have higher binding energies than the on-surface C1 and the subsurface
octahedral C3-site for all coverages $\Theta \leq 0.5$ (compare
Fig.~\ref{f2} and Table~\ref{t1}).
Thus, the first O atom adsorbed in the oxidation process binds
in a subsurface site of type B2, its binding energy being only 0.09 eV
and 0.15 eV higher (Fig.~\ref{f2}) than that of O in subsurface site
A4 and A5, respectively.
The preference for the B2 site agrees with the results of Ref.~\onlinecite{Bung97},
but we find a binding energy about 0.6 eV larger.
In addition, they find that the on-surface B2top site is much more favored
than the subsurface A4 site, in clear disagreement with our results.
These difference are likely due to the lack of Mg-atoms relaxation and the
inferior $k$-points sampling used in that study.


\begin{table}
\caption{\label{t1}
The binding energy $E_{\text{ads}}$ per O-atom for a double-site
O occupancy of different chemisorption sites (see Fig.~\ref{figsketch})
and for different surface unit cells. Two O atoms in the $4\times 4$
and $2\times 2$ surface unit cells correspond to the coverages
$\Theta =1/8$ and 1/2, respectively.
The exact vertical position of the sites B2 are distinguished by ``top"
for a position (slightly) above the top Mg-layer, and ``mix" for a
position in which some of the top Mg-atoms are above and some are
below the oxygen atom, whereas B2 denotes adsorption below all atoms
of the top-most Mg-layer. All pairs of sites are nearest neighbors.}
\begin{ruledtabular}
\begin{tabular}{lclc}
Sites &  $E_{\text{ads}}$ [eV] & Sites & $ E_{\text{ads}}$ [eV] \\
\cline{1-2} \cline{3-4}
\multicolumn{2}{c}{1/8 ML coverage}&\multicolumn{2}{c}{1/2 ML coverage}\\
\cline{1-2} \cline{3-4}
B2 \& B2    & 5.07 &  B2 \& B2    & 5.13  \\
B2 \& C3    & 4.77 &  B2mix \& C3 & 4.74  \\
B2 \& A4    & 5.03 &  B2mix \& A4 & 4.99  \\
B2 \& A5    & 5.03 &  B2 \& A5    & 5.00  \\
B2top \& B2 & 5.00 &             &        \\
\end{tabular}
\end{ruledtabular}
\end{table}


In order to check the possibility of formation of ionic
complexes (which would represent a seed for further oxide formation)
consisting of the Mg atoms of the topmost layer and subsurface and on-surface
oxygen, we calculated the most preferred sites for an additional O atom in
the presence of an already adsorbed one.
In studying possible combinations of the oxygen adsorption sites we
concentrated on the ones including the B2 sites.
As is evident from Table~\ref{t1}, even at low coverages with one O-atom
isolated from other O-atoms the addition of another O atom leads to an 
increased O-population in the B2 sites.
The two (neighboring) atoms in the B2 sites are by $\sim$0.03--0.04 eV more
favored than one atom in the B2 site and another one in either the A4
or A5 site.

\subsection{Higher oxygen coverage ($1 \leq  \Theta\leq 3$ ML)}

For 1 ML coverage with occupation in only one kind of site 
the energetic ordering of the most 
stable sites is different than for lower coverages 
(Fig.~\ref{f2} and Table~\ref{thigh}). 
The B2-type sites become less favorable than the A4 site and the actual 
B2 site is shifted to the B2top position. However, allowing for 
simultaneous occupancy of the B2 and B2top sites at 1 ML in the 
$1\times 2$ unit cell (Table~\ref{thigh}) we find that this combined 
B-site structure is by 0.06 eV more favored than the single A4 occupancy 
shown in Fig.~\ref{f2}.


\begin{table}
\caption{\label{thigh}
The binding energy $E_{\text{ads}}$ per O-atom, work function change $\Delta \Phi$, 
and the
experimental XPD reliability factor $R_{\text{MP}}$ in different chemisorption
sites for 1--3 ML oxygen coverage. For calculations involving sites within the
third Mg layer and below it (sites B7, B8, A10, and A11) one extra layer of relaxed
Mg atoms was added to the slab, which then has totally 7 layers of Mg of which the
four top-most layers are relaxed. This resulted in a small increase of the $E_{\text{ads}}$
of the order of 5 meV but has not changed the ordering of energies of the sites.
Please refer to Fig.~\protect\ref{figmediumstructures} for our naming of the 2--3 ML
structures.}
\begin{ruledtabular}
\begin{tabular}{lcrccc}
Sites & $E_{\text{ads}}$
& $\Delta \Phi$ & $R_{\text{MP}}$ & $R_{\text{MP}}$ \\
&  [eV] & [eV]  &(1.4 L)&(9.7 L)\\
\hline
\multicolumn{5}{c}{1 ML, nearest neighbors}\\
\hline
  B2 \& A5                & 5.25  & -0.64 & 0.41 & 0.30  \\
  B2mix \& A4             & 5.21  & -0.62 & 0.34 & 0.37  \\
  B2top \& C3             & 5.03  & -0.49 & 0.51 & 0.43  \\
\hline
\multicolumn{5}{c}{1 ML, one atom per unit cell}\\
\hline
  B2top \& B2             & 5.39  & -0.46 & 0.46  & 0.46  \\
  A4                      & 5.33  &  0.00 & 0.42  & 0.48  \\
  B2top                   & 5.27  & -0.99 & 0.44  & 0.46  \\
  A5                      & 5.23  &  0.11 & 0.53  & 0.42  \\
  C1                      & 5.15  & -0.12 & 0.44  & 0.46  \\
  C3                      & 5.04  &  0.62 & 0.56  & 0.51  \\
\hline
\multicolumn{5}{c}{2 ML} \\
\hline
  A4 + B7                & 5.61  & -0.07 & 0.41 & 0.50  \\
  A5 + B8                & 5.55  &  0.27 & 0.56 & 0.45  \\
  B2 + A5                & 5.54  & -0.03 & 0.44 & 0.33  \\
  Rocksalt 2ML           & 5.45  &  0.72 & 0.54 & 0.45  \\
  Flat 2ML               & 5.42  & -1.14 & 0.29 & 0.36  \\
\hline
\multicolumn{5}{c}{3 ML}\\
\hline
  A4 + B7 + A10            & 5.70  & 0.04 & 0.45 & 0.50 \\
  A5 + B8 + A11            & 5.67  & 0.09  & 0.53 & 0.43 \\
  B2 + A5 + B8             & 5.66  & 0.04  & 0.46 & 0.35 \\
  Rocksalt 3ML             & 5.61  & 0.78  & 0.55 & 0.50\\
  Flat 3ML                 & 5.47  & -1.14 & 0.30 & 0.37\\
\end{tabular}
\end{ruledtabular}
\end{table}

\begin{figure*}
\scalebox{0.75}{\includegraphics{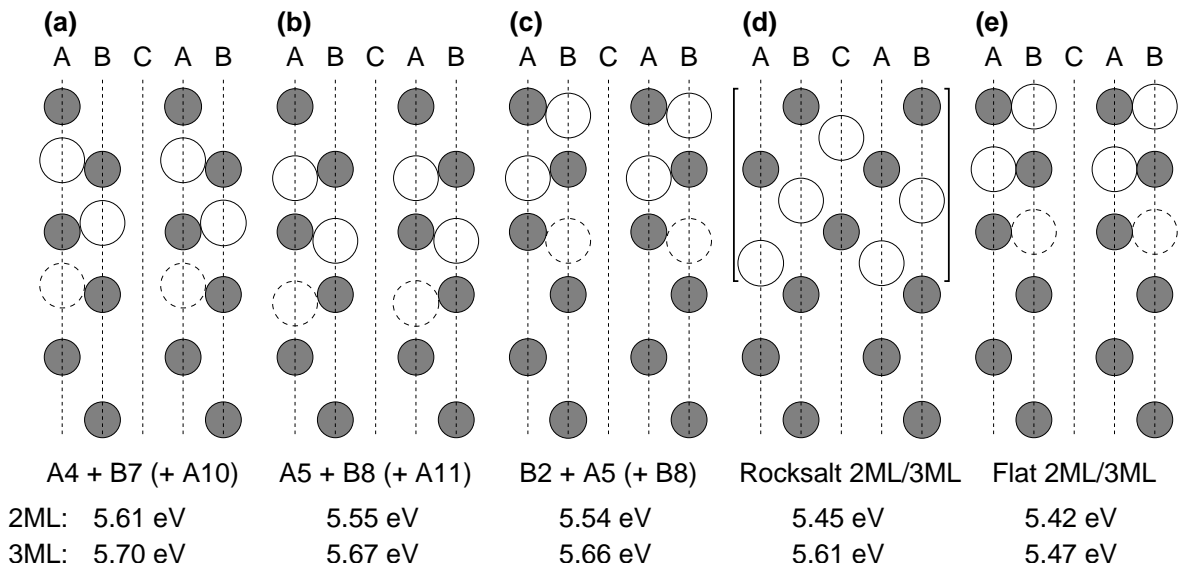}}
\caption{\label{figmediumstructures}
Schematic plot of the 2--3 ML structures, cut along the ABCAB line in 
Figure 1, and their binding energies. Small gray circles are Mg atoms, 
large white circles are O atoms. The dashed circles indicate O atoms 
that are only included in the 3 ML structure. The buckling of the Mg-O 
layers in (a)--(c) is exaggerated.
For the rocksalt structure only the 3 ML structure is shown, and the
oxide structure is in brackets to indicate that the atoms are shifted out
of the cut plane. The rocksalt structure has the atoms of the third Mg
layer, counted from the top, sitting on a bridge site of the fourth Mg layer.
The structures are in order of decreasing binding energy $E_{\text{ads}}$.}
\end{figure*}

As more oxygen is adsorbed into the Mg surface the possible adsorbate
configurations grow in number, partly due to simple combinatorics,
and partly because new adsorption sites appear within an increasingly
distorted Mg background lattice. This is already apparent
for nearest-neighbor adsorption in the low-coverage region (Table~\ref{t1})
where a slight vertical distortion of the top Mg-layer in some cases
leads to adsorption in the B2top site and in a `mixed' B2-B2top position.
In Table~\ref{thigh} we name the 1 ML configurations
by the adsorbate position, but for higher coverages
the notation is merely an approximate description of the
structure, and we refer the reader to Fig.~\ref{figmediumstructures}
for a sketch of the multitude of similar but
not identical configurations at coverages 2--3 MLs.

The stable structures found in the 2 and 3 ML coverage region fall in
two general classes: layered structures with oxygen and magnesium
in buckled or flat layers having a hexagonal structure
[Fig.~\ref{figmediumstructures} (a)--(c), and (e)], and rock-salt
structures [Fig.~\ref{figmediumstructures} (d)] on top of clean magnesium, 
strained with respect to the clean-surface lattice constant. 
The buckled structures are labeled according to the approximate positions 
of the oxygen atoms (comparing to the low-coverage adsorption sites of 
Fig.~\ref{figsketch}). The buckled structures involve subsurface oxygen only.
The vertical separation of O and Mg within a buckled layer is approximately 
0.6 {\AA}, and the layers can exist both on top of the clean Mg(0001)
[Fig.~\ref{figmediumstructures}(c)] and as subsurface layers 
[Fig.~\ref{figmediumstructures}(a) and (b)], and at 2 or 3 ML coverage.
Thus one can discriminate the structures buckled up and down, depending on
the oxygen atoms staggering over or below the Mg layer, respectively.
The flat structures [Fig.~\ref{figmediumstructures}(e)]
have vertical O-Mg separation 0.01--0.06 {\AA}, and exist for both 2 and 3 ML
coverage, but do not exist as subsurface layers. In the flat structures
the top-most oxygen atom sits slightly above the top Mg layer, corresponding
approximately to the position of site B2top in Fig.~\ref{figsketch}.
The binding energies of the 1--3 ML structures are given in Table~\ref{thigh}.
Note that for several other configurations with 2 ML or 3 ML coverage of the 
tetrahedral adsorption sites no energy minimum is found. 
This includes any subsurface flat phase, phases of mixed flat and 
buckled layers, and mixed buckle-up and buckle-down layers.
Further, we tested structures that include an octahedral (C3, C6, C9) site 
in the 2 ML and 3 ML coverage and found binding 
energies per atom at least 0.3--0.4 eV smaller than the energetically best
structures at the same coverage, similar to the results of the 
low coverage results (Fig.~\ref{f2}).

The preference for subsurface adsorption at the Mg(0001) surface was
originally suggested by a sharp work function decrease upon
oxidation.\cite{Hay81} Previous local-density-approximation DFT
calculations\cite{Bung97} have shown that only oxygen incorporation
into B2 sites lowers the work function and for one monolayer load this
lowering was found to be 0.3 eV, which is much smaller than measured
experimentally.
We do not find 1 ML adsorption in the B2 sites, but for the neighboring
B2top sites our results presented in Table \ref{thigh} confirm the lowering
of the work function. However, this lowering is three times larger than
that reported in Ref.~\onlinecite{Bung97} and we note that
our B2top sites are slightly above and nearly coplanar with the topmost
Mg layer. This larger decrease of the work function upon
oxygen adsorption agrees with recent measurements.\cite{MitOCS98}


It is generally expected that an adlayer of negative O ions will increase
the metal-surface dipole layer and the work function. For example,
this is clearly
the case for oxygen on the Al(111) surface.~\cite{KiLu01}
Our results suggest that 1 ML of on-surface (B2top) oxygen on 
the Mg(0001) surface
leads to a decrease in the work function. This decrease for on-surface O
is consistent with experimental observations.\cite{MitOCS98}
On the other hand, for 1 ML coverage the B2top sites are not the
energetically most favorable. We conjecture that 
the following scenario takes place.
At low coverage the only on-surface sites are the fcc hollows (C1),
but also the lower-lying subsurface B2 sites are exposed to O atoms arriving
at the surface. Thus an O-atom that arrives to the surface
will find the B2 site having a higher adsorption energy than
the C1 site and will immediately diffuse to the former.
The B2-sites are most favorable up to 0.5 ML coverage (Fig.~\ref{f2}).
With growing O-coverage the energetic situation changes and
the deeper tetrahedral sites become more favorable. However,
there is an energy barrier to overcome in order to get into these
lower-lying A4 and A5 sites so for 1 ML coverage, instead of 
moving further into the surface, the oxygen atoms
stay in the neighborhood of the B2 sites, now 
shifted into the B2top position. The B2top site is energetically 
slightly less favored than the tetrahedral subsurface sites
(by about 0.06 eV) but requires no crossing of significant 
energy barriers.
With an increased oxygen load (around 2 ML)
the energy barrier is lowered and some of the oxygen atoms
can move into the A5 sites. Again, the occupation of the B2top and A5
sites (the ``Flat 2ML" structure) is not the most favorable energetically
(Table~\ref{thigh})
but presumably this metastable state is most easily available for the
O atoms.
The work function change $\Delta\Phi$ for this configuration is similar
to $\Delta\Phi$ for the energetically favorable A4+B7 subsurface structure.
With a further increase of O load (3 ML) the energy barriers that
separate the lower lying sites are lowered and oxygen populates the B8 sites.
It seems that this scenario is supported by our XPD 
experiments presented in Sec.~III.

\textit{Island formation}.
In a previous theory study based on a lattice-gas-model\cite{Else01} it was 
shown that oxygen atoms that are adsorbed in the Mg(0001) surface form dense
clusters in tetragonal subsurface sites. For example, at a dosing 
corresponding to a global coverage of 1 ML, the oxygen atoms were found to 
be distributed, on average, with 55\% in site B7, 30\% in site A4, and
15\% in site A10 and with local coverages of 2--3 MLs within the clusters.

In the present, more detailed investigation we find that
past the adsorption of the first, isolated O-atoms in the oxidation process
the O-atoms show a tendency for island formation by pairing up
and showing higher binding energies in combinations with subsurface sites.
An inspection of the binding energies presented in Fig.~\ref{f2} and in
Table~\ref{t1} shows that the higher the coverage, the larger the binding
energy of the lower-lying sites (sites A4 and A5).
Our results show that in general the binding energy is
higher in  more close-packed systems, thus showing that clusters or
islands of O atoms are preferred to isolated or pairs of O atoms.
In fact, for 1 ML coverage,
having all O atoms in site A4 is preferred by a small amount to all other
subsurface sites and to the on-surface B2top site. This agrees well with the
previous lattice-gas-model study.\cite{Else01} With coverage increasing
above 0.5 ML, the O-atoms from the B2 sites are pushed up into the surface
B2top sites. Thus they end up,
at 1 ML load, in the B2top-sites located almost coplanarly with the Mg atoms
of the topmost layer, with the binding energy comparable to that of O in 
sites A4 and A5.


\begin{figure*}[t]
\scalebox{0.6}{\includegraphics{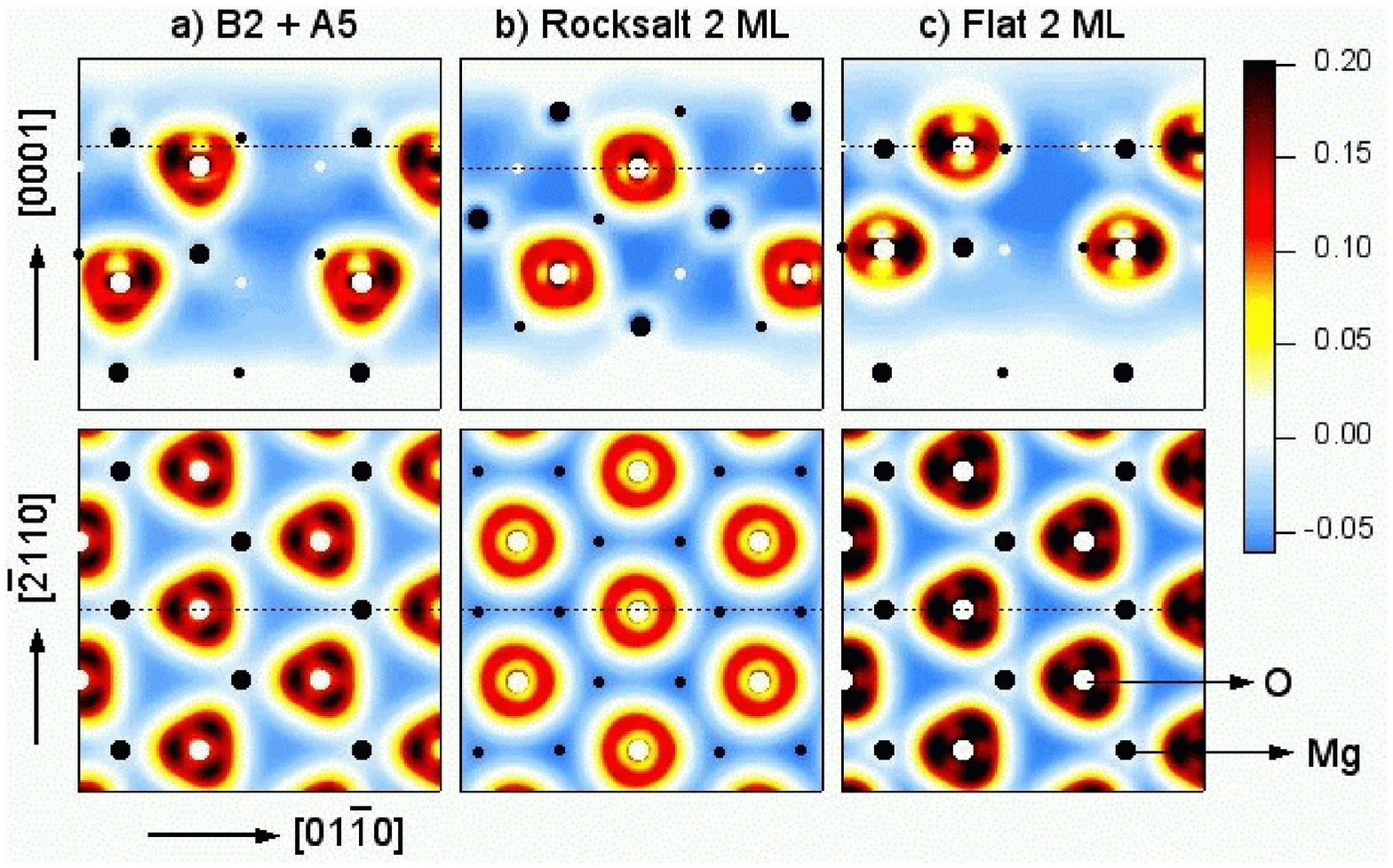}}
\caption{\label{figdens} (Color online)
The electron charge redistribution due to the presence of the O atoms.
Shown are the three 2 ML structures that have O atoms
closest to the surface, cut in the 
[$01\bar{1}0$] direction
perpendicular (top panels) and parallel to
the surface (bottom panels). Atomic positions are indicated
by white (O) and black (Mg) dots. The change in electron density
distribution is
$\Delta n(\mathbf{r})=n^{\text{Mg+O}}(\mathbf{r})
-n^{\text{Mg}}(\mathbf{r})-n^{\text{O}}(\mathbf{r})$
where $n^{\text{Mg+O}}$ is the electron density for the full system,
$n^{\text{Mg}}$ the electron density in the absence of O atoms,
and $n^{\text{O}}$ the atomic O charge density.
A negative/positive $\Delta n$ corresponds to a
depletion/accumulation of electrons (in units of $e$/\AA$^3$).
}
\end{figure*}


The preference for the formation of subsurface islands in tetrahedral
A4 sites or lower agrees both with the lattice-gas model simulations and with 
the results of Bungaro et al.~\cite{Bung97} but disagrees with
the results of Thiry et al.~\cite{Thi89} who found adsorption
in octahedral sites.
A new feature that was not observed by Bungaro et al., and which is
probably connected with a more careful treatment of the lattice relaxation
effect in our work, is the appearance of the on-surface O-islands for 1 ML
coverage, with binding energy slightly lower than in subsurface
islands of A4-type. 
Also worth noting are the relatively small binding energy
differences for oxygen in different sites for $\Theta=1$.

The tendency to form islands continues with further increase of coverage,
showing some quenching in the adsorption energy at 3 ML coverage.
A similar behavior was found in the lattice-gas model simulations.~\cite{Else01}

\textit{Layered structures}.
As the coverage is increased beyond a single monolayer the layered
structures of Fig.~\ref{figmediumstructures}(a)--(c) and (e) appear,
along with a rocksalt surface structure, Fig.~\ref{figmediumstructures}(d).
The layered structures can be described as a Mg(0001) surface
with slightly changed atomic positions to accommodate O atoms within
the top layers of the Mg surface, thereby creating a layered structure.
Each of the layers has a honeycomb structure with O
at every second vertex and Mg at the remaining vertices.  The layers
are stacked in an AA$^\prime$ stacking sequence, with O on top of Mg atoms
and \textit{vice versa} at a layer separation of approximately 2.3--2.7 {\AA}.

Again, we find (Fig.~\ref{figmediumstructures}) that subsurface
layered structures are preferred. Comparing the adsorption binding
energies listed in Fig.~\ref{figmediumstructures} it is evident that
with increased O coverage, the difference in binding energies for
the subsurface layered and rocksalt structures is diminishing, 
indicating that eventually, for still higher coverage,
the rocksalt structure may become the most favorable one.

\textit{Directional bonding}.
Figure~\ref{figdens} displays the electron charge redistribution
for three of the 2 ML structures. Whereas the rocksalt structure
(Fig.~\ref{figdens}(b)) has an almost isotropic charge redistribution
around each atom, as is typical for ionic binding, the two layered 
structures show more complicated, anisotropic charge redistributions.
The flat layered structure (Fig.~\ref{figdens}(c)) shows most
clearly the anisotropy, with more charge redistribution along the
in-plane nearest-neighbor Mg-O lines than in other directions.

\section{EXPERIMENTS}

To supplement the DFT calculations and to provide a reference
ground we have also carried out XPD measurements of the 
clean Mg(0001) surface and after low dosing of oxygen.

\subsection{Method}
XPD has been chosen because of its chemical sensitivity and its 
sensitivity to local real space order. It is a powerful technique 
for surface structural investigations,\cite{Fadley} and it has 
been shown that full-hemispherical XPD patterns provide very direct 
information about the near-surface structure. At photoelectron kinetic 
energies above about 500 eV, the strongly anisotropic scattering by 
the ion cores leads to a forward focusing of the electron flux along 
the emitter-scatterer axis. The photoelectron angular distribution, 
therefore, is to a first approximation a forward-projected image of 
the atomic structure around the photoemitters. Analysis of the 
symmetry and positions of forward-focusing maxima thus permits a very 
straightforward structural interpretation of XPD data. Furthermore, 
detailed structural parameters can be determined by comparing the 
experimental XPD patterns to calculated ones.
The relatively 
simple and efficient single-scattering cluster (SSC) formalism\cite{Fadley} 
has proven adequate in most cases. The agreement between SSC calculations 
and experimental XPD pattern can be quantified using a reliability factor 
such as the \textit{R\/}-factor $R_{\text{MP}}$ defined 
previously.\cite{RMPref,RMPexplain}

The experiments have been performed in the University of Fribourg's 
VG ESCALAB Mk II spectrometer modified for motorized sequential 
angle-scanning data acquisition. Clean Mg(0001) surfaces have been 
prepared by cycles of sputtering (500 eV Ar$^+$) and annealing 
(130~$^{\circ}$C). Before O$_2$ exposure, the surface displayed a 
sharp $(1\times 1)$ LEED pattern with little background. After O$_2$ exposure 
with the sample held at room temperature, the coverage was determined 
from the relative intensities of the O 1\textit{s\/} and Mg 2\textit{p\/} 
photoelectron peaks. Experimental O $1s$ XPD patterns were obtained 
after exposure of the Mg(0001) surface to 0.15, 0.7, 1.4 and 9.7~L of 
O$_2$, corresponding to \textit{global\/} O coverages of 0.1, 0.4, 1 and 
1.7 ML, respectively.

\subsection{Oxygen adsorbed at Mg(0001)}
The Mg(0001) surface with various coverages of oxygen atoms was
studied with LEED and XPD. Experimental XPD patterns for O$_2$ doses
of 0.15 and 9.7~L
are shown in Fig.~\ref{figexpSSC}. The most prominent 
feature of the 0.15 L pattern is a strong intensity maximum at normal 
emission. This observation is a clear and direct evidence for O adsorption 
below the topmost Mg layer: Only an atom located directly above the oxygen 
photoelectron emitter can give rise to a forward-focusing maximum at 
normal emission, i.e., in the center of the plot. A further conclusion can 
immediately be drawn from the evolution of the XPD pattern with O$_2$ 
exposure. Apart from differences due to counting statistics, the experimental 
XPD patterns are strikingly similar, not only regarding the intensity maximum 
at normal emission but regarding all the prominent diffraction features. 
It must therefore be concluded that the local atomic geometry is the same 
over the entire exposure range.

The same conclusion is obtained from the evolution of the LEED pattern with 
increasing O$_2$ exposure: The LEED pattern stays ($1 \times 1$), and only 
the background rises slightly, which indicates an increasing amount of disorder 
with increasing coverage. 
Since no superstructures are observed, the presence of any other islands than
islands of ($1 \times 1$) periodicity that are larger than about 
50--100 \AA\ can therefore be excluded.

In other words, both the LEED and XPD measurements indicate that locally the 
geometry is the same irrespective of the global coverage (at least below 1.7 ML). 
Already at 0.1 ML global coverage this local structure thus contains at 
least 1 ML oxygen, and with increasing coverage it will grow in domain 
size without significant changes in geometrical structure. Therefore, 
determining the local geometrical structure at 1ML local coverage will 
determine the geometries even at the lowest global coverages accessible 
to the experiments.

In the further analysis we compare simulated SSC diffraction pattern to 
experimental XPD images of the Mg(0001) surface both with dosing 1.4 L 
(approx.~1 ML global coverage) and 9.7~L (approx.~1.7 ML global coverage).

\section{Comparison between experiment and theory}


\begin{figure}
\scalebox{0.4}{\includegraphics{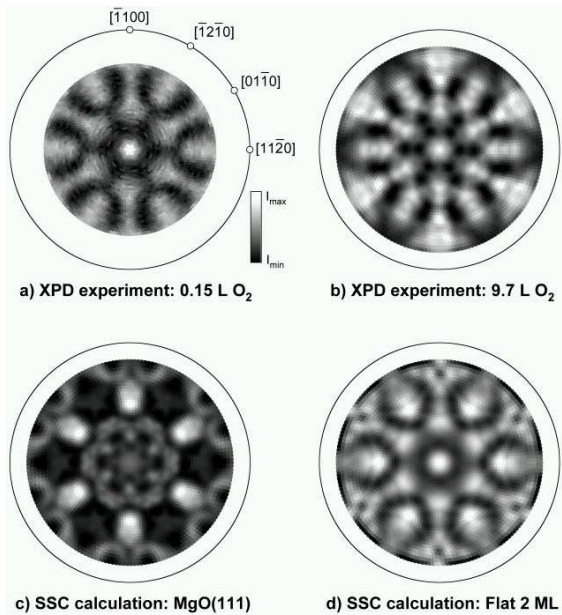}}
\caption{\label{figexpSSC}
Experimental O $1s$ XPD patterns after exposures of the Mg(0001) 
surface to 
(a) 0.15~L and 
(b) 9.7~L of oxygen. 
(c) SSC calculation for the MgO(111) rocksalt structure. 
(d) SSC calculation for the 2 ML layered oxide structure 
(``Flat 2ML'') yielding best agreement with the experimental 
1.4~L XPD pattern.} 
\end{figure}

That the O/Mg(0001) system does not simply form a MgO(111) rocksalt 
structure is clearly seen by comparing a simulated diffraction pattern 
for MgO(111) [Fig.~\ref{figexpSSC}(c)] to the experimental 9.7~L XPD 
pattern shown in Fig.~\ref{figexpSSC}(b): The simulation does not 
reproduce any of the diffraction features satisfactorily. 
Therefore, the most relevant atomic structures, as determined 
by the DFT calculations, 
were used to simulate XPD patterns by means of SSC calculations. The 
resulting SSC calculations were compared to the experimental 
1.4~L and 9.7~L XPD patterns, 
the latter of which is shown in Fig.~\ref{figexpSSC}(b).

None of the SSC calculations for an oxygen atom in an isolated site gave 
satisfactory agreement with experiment. The calculations, however, confirm 
the conclusion drawn from the experimental patterns that one oxygen atom 
is adsorbed directly below a Mg atom: Only the simulations for oxygen in 
the A4 and A5 sites exhibit the characteristic intensity maximum at 
normal emission that is seen in experiment [Fig.~\ref{figexpSSC}(a),(b)]. 
The fact that the experimental 
XPD patterns do not agree with simulated diffraction patterns of disperse, 
low local coverages even for low overall coverage in the experiment is 
consistent with the LEED measurements discussed above, indicating again 
that the oxygen will form island with a local coverage of 1 ML or more. 
Accordingly, further SSC calculations concentrated on the atomic positions 
obtained from DFT calculations using a $1\times 1$ surface cell.

Among the simulated diffraction patterns for 1 ML local coverage only the 
site A4 and the site A5 SSC calculations exhibit a central maximum.
In contrast, the SSC calculations for each 
of the C3, B2top, and C1 sites (leaving all other sites empty) do not 
show such maximum. This means that the comparison between SSC calculations 
and experimental XPD patterns thus disqualifies the site B2top-only 
occupation at 1 ML local coverage, in agreement with the theoretical binding 
energy being smaller for B2top than for A4 at 1 ML coverage. Along with 
the binding energy obtained from the DFT calculations the comparison to 
the XPD patterns also rules out octahedral subsurface (site C3) occupation, 
in disagreement with the results of Thiry et al.~\cite{Thi89}

Consequently, SSC calculations were also performed for the atomic 
positions obtained from DFT calculations considering 2 and 3 ML coverage.
In Table~\ref{thigh} we summarize the 1 ML, 2 ML and 3 ML local coverage 
SSC simulations [one to three O atoms per $(1\times 1)$ cell] by listing 
their reliability factors with respect to the 1.4~L and 9.7~L exposure 
XPD data. Best agreements between experiment and SSC calculations (lowest 
reliability factors) are found for the layered structures ``B2mix \& A4'' 
and ``B2 \& A5'' at 1 ML local coverage, ``Flat 2ML'' 
[Fig.~\ref{figexpSSC}(d)] and ``B2 + A5'' 
at 2 ML local coverage, and ``Flat 3ML'' and ``B2 + A5 + B8'' for 3 ML
local coverage. These results are consistent with the work function changes 
discussed above.

The general trend is that the rocksalt structures (2 ML and 3 ML) do not 
agree very well with the experimental XPD patterns (high values of 
$R_{\text{MP}}$), whereas some of the layered structures agree better. 
In particular we notice that among the 2 ML structures the flat 
structure (``Flat 2ML'') has the best agreement with the 1.4~L experiment, 
whereas one of the buckled layer structures (``B2 + A5'') fits best to 
the 9.7~L experiment. This indicates that as the global coverage is 
increased by higher O$_2$ dosing, more of the surface becomes covered 
with buckled Mg and O layers and less with the flat surface. 

For the 1.4~L dosing experiment the flat surface is favored, in disagreement
with the DFT results.
1.4~L dosing corresponds to $\sim$1 ML of global coverage, but since the 
distribution of oxygen is nonuniform due to the tendency of the atoms to 
form islands the experimentally observed 
surface is patched.
For small surface oxide patches of 2 MLs (or more) local oxygen coverage 
the strain can get 
released over most of the island and thus, experimentally,
mostly the flat structure is seen at low dosing.
This gives a low $R_{\text{MP}}$-value for the flat structure, but still 
keeps the order of the buckled and rocksalt structure found in DFT
calculations: buckled is also at 1.4 L more favored experimentally 
than rocksalt.
Thus, the relevant dosing (among the measured ones) to use for 
comparing DFT and the 
XPD experimental structures is 9.7~L. This structure probably does not 
fully cover the surface (if clusters are at least 2 ML dense), 
but certainly does so to a larger extent than for dosing 1.4~L. 
At dosing 9.7~L the DFT calculations and the XPD experiment agree on 
the buckled surface. 

Based on both the calculated and experimental structures discussed above 
we conclude that at relatively low dosage (corresponding to 2--3 ML coverage) 
O/Mg(0001) forms the layered oxide structure. 
The rocksalt structure typical of MgO starts to grow only at higher 
O dosing.

\section{Summary and CONCLUSIONS}

We have performed extended first principles calculations 
of oxygen adsorption and of the initial stages of Mg(0001) oxidation. 
A variety of configurations and a wide range of coverages were considered 
in order to determine the most stable structures. These were compared 
with x-ray photoelectron diffraction experiment and simulations.
At low coverages ($\Theta\leq 0.5$ ML) both our DFT calculations and
experiment show that oxygen adsorbs in subsurface sites.
Our DFT calculations show that the first O atom chemisorbed in 
the oxidation process binds in a subsurface
tetrahedral site of the B2 type. The importance of substrate lattice 
relaxation in accurate determination of the most stable sites is demonstrated.
At higher coverage O adsorbed in subsurface tetrahedral sites shows a 
tendency to
form subsurface islands which results in an increased binding energy. 
For the 2 ML (3 ML) coverages we find some rather unanticipated surface 
oxide structures, consisting of two (three) mixed oxygen-magnesium layers 
on top of an almost undistorted Mg(0001) surface.
These layered oxide structures have hexagonal symmetry and can be 
flat or buckled. 
For 2--3 ML coverage the rocksalt structure is found to be unfavorable 
compared to the buckled layer structures. However, our DFT results show 
that the rocksalt structure may become energetically competitive at
an increased coverage.

\begin{acknowledgments}
This work was supported in part by the Swedish Foundation for Strategic
Research (SSF), the Swedish Research Council (VR), The Swedish Foundation
for International Cooperation in Research and Higher Education (STINT),
the Carl Tryggers Foundation, and the Polish State Committee for Scientific
Research (KBN), project 5 P03B 066 21.
R.F.\ would like to thank P.\ Aebi for continuous support and help with the
XPD experiments. The allocation of computer time at the UNICC facility at
Chalmers and G{\"o}teborg University is gratefully acknowledged.
\end{acknowledgments}


\end{document}